\documentclass[letter]{aa} 

\usepackage[normalem]{ulem}

\usepackage{graphicx}
\usepackage{txfonts}
\usepackage{xcolor}
\usepackage{soul}

\newcommand{\kms}{{\mathrm{\, km \,s^{-1}}}}

\usepackage[colorlinks=true,linkcolor=blue,citecolor=blue,urlcolor=blue]{hyperref}

\newcommand{\cair} {\ion{Ca}{ii}~8542~\AA\xspace}

\newcommand{\cak} {\ion{Ca}{ii}~K\xspace}

\def\specchar#1{{\sc{#1}}}  
\def\Halpha{\mbox{H\hspace{0.1ex}$\alpha$}\xspace}

\def\CaII{\mbox{Ca\,\specchar{ii}}\xspace}

\def\MgII{\mbox{Mg\,\specchar{ii}}\xspace}

\def\Lw{\textit{Lightweaver}}
\def\Pw{\textit{Promweaver}}

\begin{document}

\title{Spectral Characteristics of a Rotating Solar Prominence in Multiple Wavelengths}

   \author{ 
          A.G.M. Pietrow\inst{1}\inst{2}
          \and
         V. Liakh\inst{1}
          \and
          C.M.J. Osborne\inst{3}
          \and
          J. Jenkins\inst{1}
          \and
          R. Keppens\inst{1}
          }

   \institute{
   Centre for mathematical Plasma Astrophysics, KU Leuven, Celestijnenlaan 200B, B-3001 Leuven, Belgium
   \and
   Leibniz-Institut für Astrophysik Potsdam (AIP), An der Sternwarte 16, 14482 Potsdam, Germany
   \and
   SUPA School of Physics and Astronomy, University of Glasgow, Glasgow, G12 8QQ, UK
   \\
       \email{apietrow@aip.de}
   }

\date{Draft: compiled on \today}

\abstract{
We present synthetic spectra corresponding to a 2.5D magnetohydrodynamical simulation of a rotating prominence in the \cair, \Halpha, \cak, \MgII{ k}, Ly $\alpha$, and Ly $\beta$ lines. The prominence rotation resulted from angular momentum conservation within a flux rope where asymmetric heating imposed a net rotation prior to the thermal-instability driven condensation phase. The spectra were created using a library built on the \Lw{} framework called \Pw{}, which provides boundary conditions for incorporating the limb-darkened irradiation of the solar disk on isolated structures such as prominences. Our spectra show distinctive rotational signatures for the \MgII{ k}, Ly $\alpha$, and Ly $\beta$ lines, even in the presence of complex, turbulent solar atmospheric conditions. However, these signals are hardly detectable for the \cair, \Halpha, \cak spectral lines. Most notably we find only a very faint rotational signal in the \Halpha line, thus reigniting the discussion on the existence of sustained rotation in prominences. 
}

 \maketitle
%

\section{Introduction}
Prominences are cold and dense coronal structures that appear in positive contrast against their background when protruding above the solar limb, and in a negative contrast when seen on-disk (instead termed `filaments').
In both cases, complex magnetic fields such as flux ropes (FR) and sheared arcades (SA) are responsible for the suspension of these structures \citep{Mackay2010, Terradas2016}. 
 Due to their suspension within the solar corona, prominences are commonly considered to be subject to low plasma-$\beta$ dynamics, which means that the plasma motions are dominated by magnetic tension and magnetic pressure dynamics \citep[e.g.][]{Jenkins2021,Brughmans2022,Jercic2024}.
Such bounding to the local magnetic field strongly indicates that the highly dynamic nature of prominence plasma thus maps directly to the underlying field topology since the magnetic field acts to guide the trajectory of the associated plasma.
A commonly studied example of such field-aligned dynamics is rotation, which was first described by \citet{secchi1875} as `spiral structures', later by \citet{Young1896} as `whirling waterspouts', but is primarily attributed to \citet{Pettit25} under the nomenclature of `prominence tornadoes'. 

Since then, these events have been reported more frequently. Numerous instances have been studied in the feet of prominences \citep{Su:2012apjl,Su:2014apjl,Orozco:2012apjl,Yan:2013aj,Yan:2014apj,Wedemeyer:2013apj,Wedemeyer:2014pasj} as well as within coronal cavities \citep{Ohman:1969solphys,Liggett:1984solphys,Schmit:2009apjl,Wang:2010apjl,Li:2012apjl,Mishra:2020solphys}, with estimated rotational velocities ranging from $5$ to $75\kms$.
The events observed within cavities are thought to be plasma moving along a helical FR, which appears as rotational motion when observed along the axis of the helix \citep{Li:2012apjl,Panasenco:2014solphys}. Over time, these flows along a helical FR may lead to plasma drainage down to one of the FR footpoints \citep{Wang:2010apjl,Wang:2017apj}. 

Unlike the more unequivocal rotational motions observed in erupting flux ropes as they unwind, the sustained ($>1$h) rotations of prominences has recently been called into doubt by \citet{Levens2018} and \citet{Gunnar23}, who suggest that such motions can also be explained as oscillations, counter-streaming flows, and non-rotational 3D movements projected to seem like helical motions.

Over the last decade, 2D and 3D magnetohydrodynamic (MHD) simulations of the formation and evolution of solar prominences have improved dramatically to where current simulations \citep[e.g.,][]{Jenkins2021,Jenkins2022,Donne2024} exceed the resolution of most observations and are comparable to the new Daniel K. Inouye Solar Telescope  \citep[][]{Rast2021}.

Recently, \citet{Liakh23} presented the self-consistent emergence of sustained rotation in plasma motions within a cavity, using a 2.5D MHD simulation of the levitation-condensation scenario of the prominence formation in a non-adiabatic, gravitationally stratified coronal volume, accounting for the effects of turbulent heating at the footpoints. The flux-rope embedded prominence forms due to thermal instability as lower-lying coronal matter gets levitated during the flux-rope formation, and the contracting condensation inevitably starts rotating.

Simulated Atmospheric Imaging Assembly \citep[AIA, ][]{Lemmen2012} proxies \citep[e.g.,][]{Xia2014,VanDoorsselaere2016,Gibson2016}, and synthetic spectra act as a link between simulations and observations. While the former has been widely employed over the last decade, the latter has only recently come into fashion for those more detailed prominence models. The cause is two-fold: the recent advent of self-consistent prominence formation and evolution models as well as the relative difficulty of synthesizing spectral lines that form outside of local thermodynamic equilibrium (LTE) conditions \citep[e.g.][]{Jenkins2023,Jenkins2024}. Only recently have full multi-dimensional non-LTE transfer computations on such highly resolved prominence models become a possibility \citep{Osborne2024}.

\begin{figure*}
\centering
\includegraphics[width=0.9\textwidth]{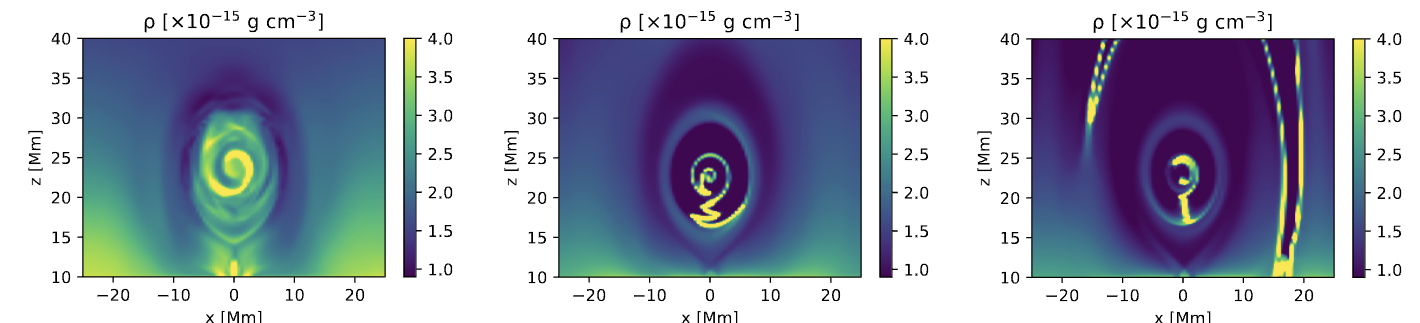}
    \caption{The density distribution in the central part of the numerical domain at 42, 80, and 142 minutes, showing the different stages of evolution of the prominence as well as the longevity of the rotational motion. An animated version of this figure is available online (see movie~1). }
    \label{fig:simulation}
\end{figure*}

In this work we use \Pw{}\footnote{\href{https://zenodo.org/records/11358632}{Promweaver: https://zenodo.org/records/11358632}}, a library built on the \Lw{} framework \citep{Osborne2021} to forward model the simulation presented in \citet{Liakh23}. With this, we will for the first time offer spectral information on how rotating structures such as the one described in this work would appear if observed, and whether or not the rotational signal is clear enough to distinguish in a turbulent solar atmosphere. 

\section{Simulations with MPI-AMRVAC}\label{sim}

We use the 2.5D numerical simulation presented in \citet{Liakh23} showing a systematically rotating prominence within its coronal cavity which is driven in a way similar to the one described in \citet{Wang2010}. While this paper provides a brief overview of the numerical model and its evolution, a comprehensive description is available in \citet{Liakh23}.

The numerical experiment was performed using the open source, adaptive-grid, parallelized Adaptive Mesh Refinement
Versatile Advection Code\footnote{  \url{http://amrvac.org}} \citep[MPI-AMRVAC, ][]{porth2014, xia2018, keppens2020, keppens2023}. The simulation box has a domain of 48 x 144 Mm (65 x 195") with adaptive mesh refinement achieving a minimum grid cell size of 31.25 km (0.04"). The initial magnetic field configuration consists of a force-free sheared arcade, as described by \citet{Jenkins2021}. Converging and shearing motions at the bottom boundary (for the $V_x$ and $V_z$ velocities) were implemented to form the FR according to \citet{vanBallegooijen:1989apj}.

To simulate turbulent heating from the lower solar atmosphere, randomized heating is included at the bottom of the numerical domain, as in \citet{Zhou2020,Li:2022apj,Jercic2024}. This heating significantly influences the FR formation, introducing asymmetry in temperature and density distributions with respect to the polarity inversion line (PIL). This leads to a coherent flow along magnetic loops, which evolves into a rotational pattern as a twisted FR forms. The plasma near the FR center continues rotating for over an hour, initially reaching speeds of over $60\kms$. Prominence plasma located farther from the FR center forms at the apex, drains down, and oscillates around the bottom of magnetic dips with a velocity amplitude of approximately $30\kms$. This is demonstrated by means of synthesized AIA images of the 304, 211, 193, and 171~\AA\ channels \citep[Fig.~4 of][]{Liakh23} showing the simultaneous rotation of both coronal plasma and condensations as well as the formation of the coronal cavity and dimming region in the overlying loops. 

The simulation runs for 1000 time steps with a $\sim$~8.59~s cadence, spanning a total of 142 minutes. In Fig.~\ref{fig:simulation} the density distribution at the start, middle, and end of the simulation is shown, with the rotation being illustrated in movie~1. The condensation tail forms later on and remains oscillating until the final stage of the numerical experiment. 

The MPI-AMRVAC data solved on an AMR grid were converted into uniformly spaced NetCDF \citep{Rew1990} files using the yt-project \citep{Turk2011}.

\section{Spectral synthesis with \Pw{}}\label{sim}
\begin{figure*}
\centering
\includegraphics[width=0.9\textwidth, trim={0cm 0.4cm 0 0.3cm},clip]{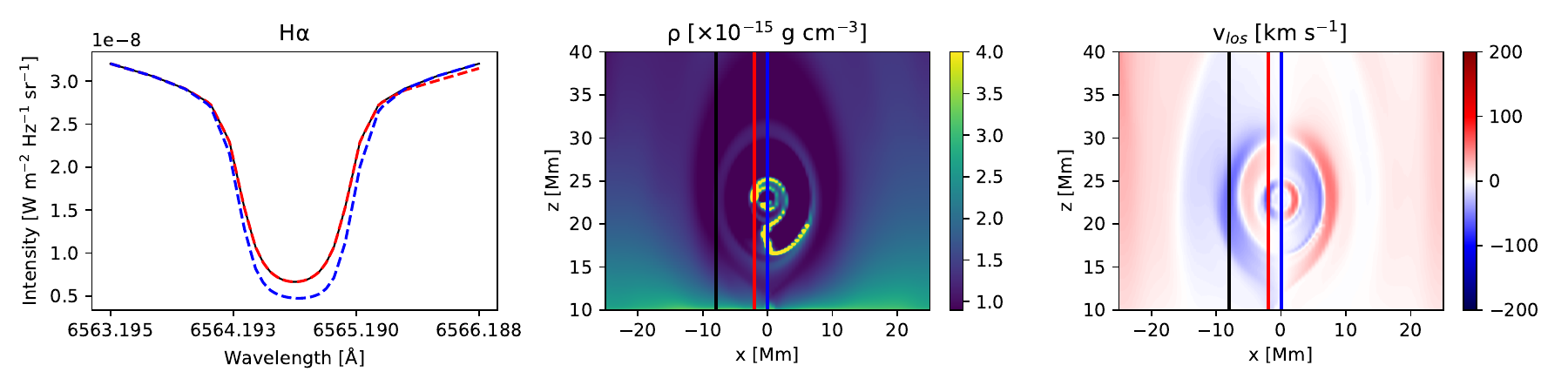}
\includegraphics[width=0.9\textwidth, trim={0cm 0.4cm 0 0.3cm},clip]{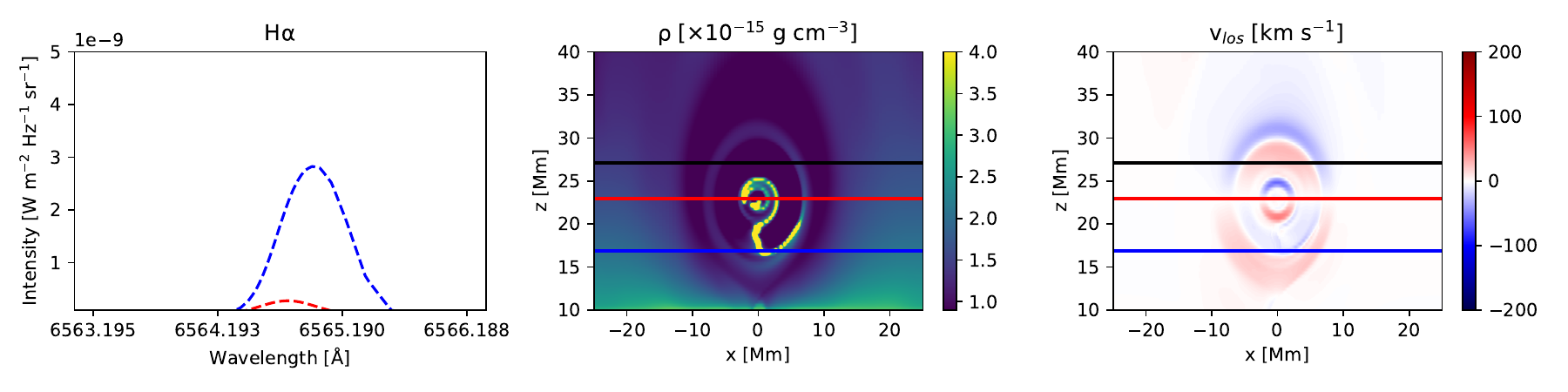}
    \caption{Example synthesized spectra along three rays at t=85 min. A ray representing the background in black, rotating condensation in red, oscillating tail in blue. The top row shows the spectra resulting from filament synthesis, with the three rays marked in the line-of-sight velocity and density map. Here the black line overlaps with the red line everywhere except at the red wing. The bottom row shows the same but for prominence synthesis, with the black line being close to zero across the spectrum. An animated version of both views is available online (see movie~2 and movie~3).}
    \label{fig:spectralcuts}
\end{figure*}

\Pw{} is an extension to the \Lw{} framework designed for modeling isolated structures within the solar atmosphere.
As of v0.3.0 \citep{Promweaver030} it provides simplified classes for the boundary conditions as described in \citet{Jenkins2023} and for employing these with stratified atmospheres (such as columns from MPI-AMRVAC) whilst optionally conserving charge and pressure \citep{Jercic2024}.
The boundary conditions employed in \Pw{} sample the limb-darkened radiation emerging from the solar disk arriving at the prominence and can self-consistently account for Doppler shifts relative to the disk  \citep[see also ][]{Peat2024}.
We adopt a semi-empirical FAL-C \citep{Fontenla1993} as a reference quiet-sun model, from which the limb-darkened radiation across the spectral range of interest is tabulated, via a plane-parallel treatment and serves as input to the boundary conditions.
This library also contains classes for reproducing the PROM code series of models \citep[e.g.][]{GHV1993, HGV1994} for both the isothermal and isobaric cases, as well as those with a prominence corona transition region \citep[PCTR, as described in][]{Labrosse2004}.

The \Lw{} framework \citep{Osborne2021} employs the multi-level accelerated lambda iteration approach of \citet{Rybicki1992} with the iterative procedure for determining partial frequency redistribution (PRD) of \citet{Uitenbroek2001} and the hybrid PRD approximation for moving atmospheres as described in \citet[][]{Leenaarts2012}. We use the standard settings for the formal solver and interpolation techniques.
In this work, we model and investigate spectral lines of H, \CaII{}, and \MgII{}.
Partial frequency redistribution is employed for the resonance lines investigated for all of these species, namely: Lyman $\alpha$ \& $\beta$, \CaII{} K \& 8542~\AA, and \MgII{} k. Similarly to \citep{Jenkins2023} we find that the signal produced by the other two lines of the \CaII{} IR triplet, as well as \CaII{} H and \MgII{} h are nearly identical to their counterparts, and we thus leave them out of our analysis.

\section{Results, and Discussion}

In this study, we consider two axis-aligned projections, a top-down view which we dub `filament view', and a sideways `prominence view'. This is illustrated in the second and third columns of Fig.~\ref{fig:spectralcuts} where vertical and horizontal rays can be seen for filament and prominence view respectively. The first column shows the synthetic \Halpha profiles obtained from the synthesizing of these rays together with the appropriate FAL-C boundary conditions. The second column shows the density map in the same units as Fig.~\ref{fig:simulation}. The third column shows the line-of-sight velocity (e.g. along the integration direction). An animated version of both figures is available online (see movie~2 and movie~3).

The black line in both views represents a `quiet' reference profile, which is traced through a part of the simulation where the prominence density is too low to affect the underlying \mbox{FAL-C} atmosphere. The red line crosses a high-density rotating kernel which manifests itself in both views as a redshifted signal of approximately 60~km~s$^{-1}$. The blue line crosses the high-density oscillating condensation tail below the rotating part of the prominence. The much lower velocities of this region do not noticeably shift the profile, but the increased column mass results in opacity broadening of the line as well as a saturation of the line core in filament view. This is in line with \citet{Leenaarts2012}, where the authors show that the \Halpha line opacity is mainly sensitive to mass density under chromospheric conditions. These resulting spectra are also similar to the observations shown in Fig.~6 of \citet{Schwartz2019} and Fig.~2 of \citet{Kuckein2016}, showing that this tail is similar to typical prominence plasma.

\begin{figure*}
\includegraphics[width=0.95\textwidth, trim={0cm 0 0 1cm},clip]{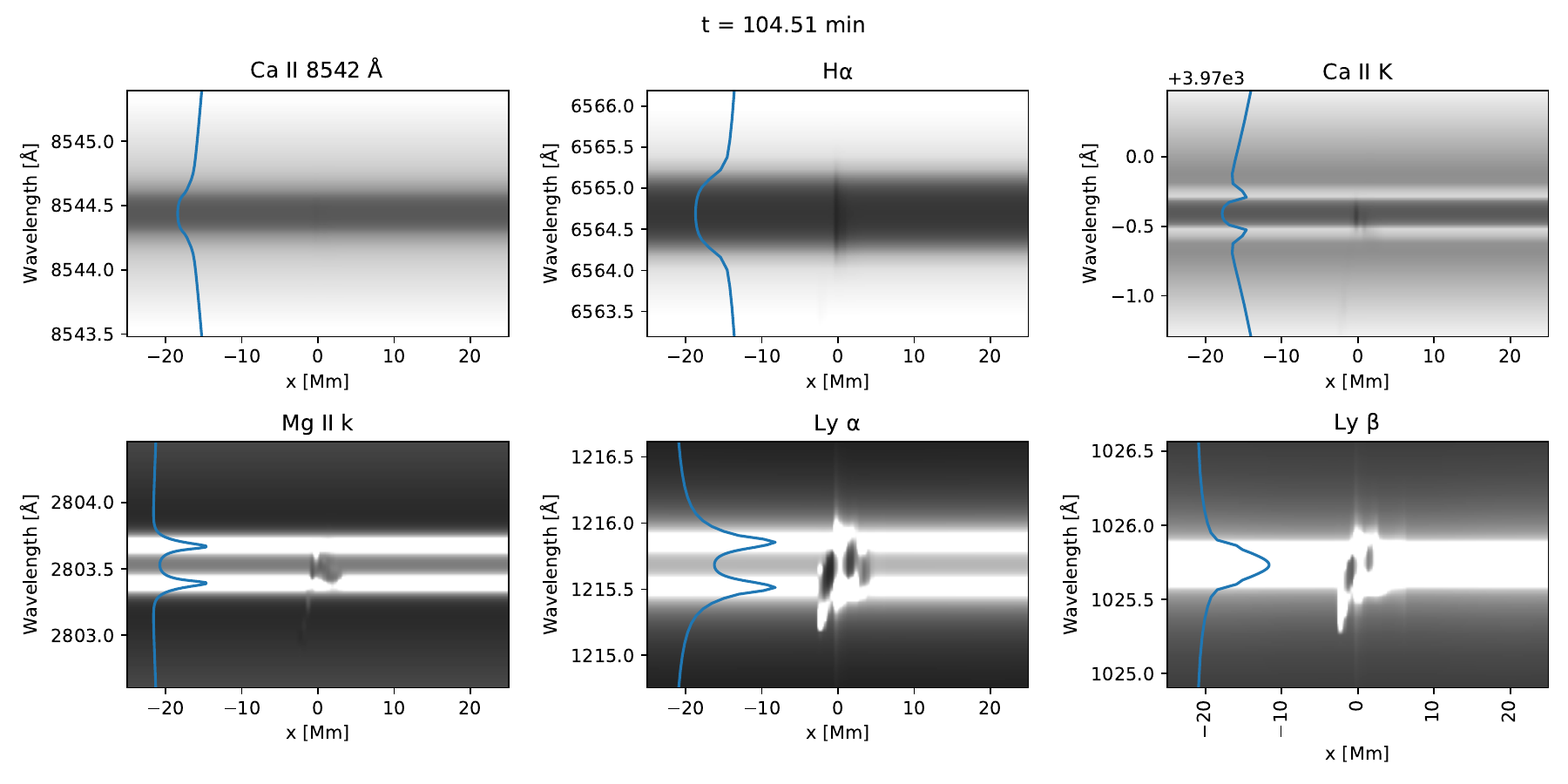}
    \caption{Simulated spectral slit across the filament view at t=85~min (See first row of Fig.~\ref{fig:spectralcuts}.), showing the \cair, \Halpha, \cak, \MgII{ k}, Ly $\alpha$, and Ly $\beta$ lines. An animated version of this figure is available online in movie~4.}
    \label{fig:specfil}
\end{figure*}

\begin{figure*}
\includegraphics[width=0.95\textwidth, trim={0cm 0 0 1cm},clip]{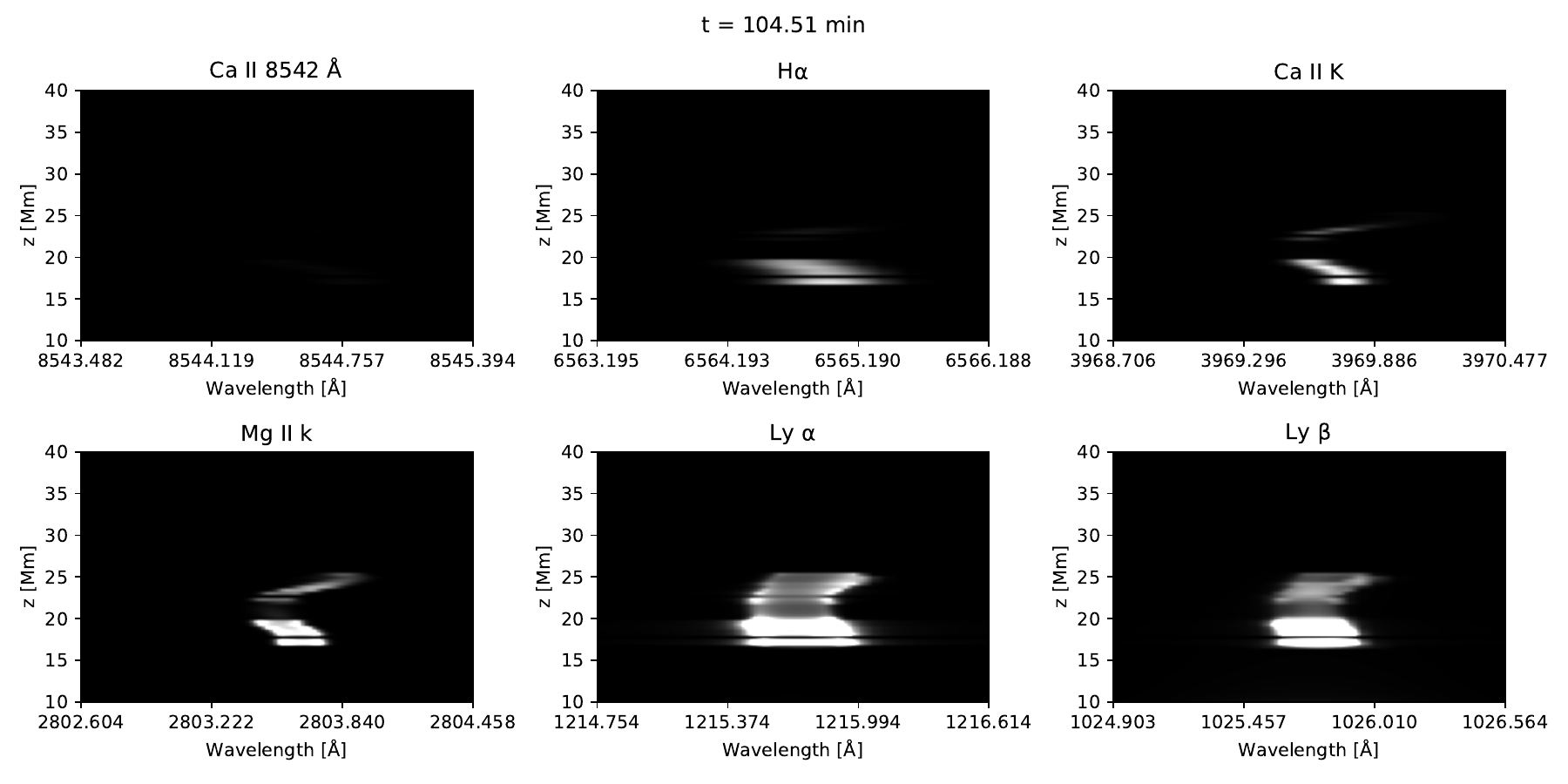}
    \caption{Same as Fig.~\ref{fig:specfil} but for the prominence view. An animated version of this figure is available online in movie~5.}
    \label{fig:specprom}
\end{figure*}

We repeated the synthesis steps for both views across the entire time series. However, due to the relatively high computational cost of creating these spectra, we opted to sample the data sparsely in both the temporal and spatial directions. Temporally each 10$^{th}$ snapshot was synthesized, resulting in 100 snapshots spanning the full 142-minute  duration of the simulation with an 86~s cadence. Spatially, each snapshot was sampled at each 16$^{th}$ column, resulting in 100 columns across the 50 Mm domain thus giving an effective resolution of 0.5~Mm or 0.68". Both numbers are within an order of magnitude of current generation ground and space-based telescopes \citep[e.g. ][]{Scharmer03,Pontieu2014,Li2022}. On a 2022 MacBook Air, a single snapshot at this sparse resolution takes an average of six hours to complete. 
Currently, we do not exploit the sparsity of the model as large fractions of many columns have temperatures in the MK regime and are not meaningfully involved in the radiative transfer calculation but still consume memory and processing power.

In Figs. \ref{fig:specfil} and \ref{fig:specprom} we show a simulated spectral slit across the line-of-sight of the filament and prominence respectively in the \cair, \Halpha, \cak, \MgII{ k}, Ly $\alpha$, and Ly $\beta$ lines, ordered from low to high disk formation heights \citep[e.g. ][Fig. 1]{Jaime19}. Both slits correspond to the atmosphere of the same snapshot as shown in Fig.~\ref{fig:spectralcuts} at 85~minutes into the simulation, with each ray in the spectral direction corresponding to one synthesized spectrum. In both cases, we find that only the high-density rotating kernel and condensed oscillating tail affect the underlying spectrum. The evolution of the entire time series can be viewed in movie~4 and movie~5. 

\subsection{Filament view}
In the filament view, hardly any signal can be perceived in \cair throughout the time series. In \Halpha and \cak we are primarily sensitive to the higher-density condensation tail which manifests itself as an absorption feature in the line core and primarily moves around in the spatial direction showing very little Doppler shift. The rotating kernel manifests itself faintly as a highly ($\sim$~60~km s$^{-1}$) Doppler-shifted absorption feature that alternates between red and blueshift, much like the red line in Fig.~\ref{fig:spectralcuts}. The signal of the rotating kernel seems most distinct in \MgII{ k}, where a stronger absorption feature can be seen stretching across the spectral dimension, while in the Lyman lines the signals of the two features are strongly overlapping. 

To our knowledge, there are no studies focusing on the evolution of filament spectra at a single location, thus a direct comparison with observation is difficult. However, the synthetic spectra have shifts of the same order of magnitude as the erupting filament observed by \citet{Madjarska2022}.

\subsection{Prominence view}
In the prominence view the signal is scaled to a tenth of the continuum for the top row, a quarter of the emission peak for \MgII{ k}, and to the emission peak in the Lyman lines. Here we find a similar result to that of the filament view, where the rotating feature is best seen in the magnesium and hydrogen resonance lines, while barely being visible in the two lower-forming lines. Likewise, two Doppler patterns are visible; the high-density condensation tail between 15 and 20 Mm, and the rotating kernel between 20 and 30 Mm. The former shows what seems to be a broadening, but in fact, is the `back-and-forth' motion of the tail. The rotating component produces a diagonal component on the slit which exhibits the strongest Doppler shifts when the rotating kernel is in the lower left or upper right position, moving primarily in the direction of the line of sight. The rotational signal is once again best seen in the \MgII{ k} line. 

The signature left by the rotating kernel has a close resemblance to the observations shown in Fig.~8 of \citet{Barczynski2023} and to a lesser degree the spectra shown in Fig.~7 of \citet{Barczynski2021} which more closely resemble the oscillating tail. 

\subsection{Model Dimensionality}

An important limitation of this work is the dimensionality of the model used to synthesize these results, as 2.5D simulations of prominence formation and evolution often underestimate the total prominence mass due to the limited plasma reservoir available. In contrast, a 3D model allows for more realistic plasma evaporation driven by randomized heating at the footpoints of the anchored FR. The resulting increase in total prominence mass has significant implications for spectral synthesis, especially for optically thick lines like \Halpha, which are highly sensitive to density variations. Additionally, as discussed in \citet{Liakh23}, the 2.5D setup cannot account for certain key physical processes, such as field-aligned pressure gradients, leading to a lack of physical attenuation of motions. Finally, in a 3D FR model, rotational flows will eventually result in plasma drainage at one of the footpoints, a process important for understanding the lifecycle and stability of prominences. Future work can build on recent full 3D models of realistic prominences as simulated by \citep{Donne2024}.

\section{Conclusions}\label{conclusions}

Following the procedure laid out in \citet{Jenkins2023} we use \Pw{}, a library built on \Lw{}, to create synthetic spectra of the 2.5D MPI-AMRVAC simulation presented in \citet{Liakh23} in a sideways (prominence) and top-down (filament) view.

The simulation was sampled sparsely at a resolution of 0.68" (0.5~Mm) and a cadence of 86~s spanning 142 minutes, for the 65 x 195" (48 x 144 Mm) simulation box. The resulting spectra of the \cair, \Halpha, \cak, \MgII{ k}, Ly $\alpha$, and Ly $\beta$ lines are shown for one snapshot in Figs. \ref{fig:specfil} and \ref{fig:specprom}, and the full time series can be viewed in movie 4 and movie 5. 

In both views we find a Doppler signature left by the continuously rotating kernel of up to 60~km~s$^{-1}$, however in both views the visibility of the effect is negligible in \cair, and barely detectable in \Halpha and \cak. The \MgII{ k} line gives the clearest signal while the Lyman lines show a blended response between the condensed oscillating tail and the rotating kernel which may be harder to disentangle for slower rotations.

Our current findings suggest that the nature of rotating prominence makes their rotational component challenging to detect in classically used lines such as \Halpha, although this may improve once a 3D prominence model is created and studied in the same way. However, this could explain why no such effects were found by \citet{Levens2018} and \citet{Gunnar23}.
For this reason, we suggest that future observational studies focus on Interface Region Spectrograph \citep[IRIS ][]{Pontieu2014} observations where sparse, high signal-to-noise \MgII{} rasters are used to hunt for such signals in both filaments and prominences.
%
%

\begin{acknowledgements}
  RK and AP received funding from the European Research Council (ERC) under the European Union Horizon 2020 research and innovation program (grant agreement No. 833251 PROMINENT ERC-ADG 2018) and RK is supported by FWO projects G0B4521N and G0B9923N.
CMJO is grateful for the support of a Royal Astronomical Society fellowship.
We acknowledge productive discussions with Dr Veronika Jerčić and Dr Malcolm Druett, and thank the anonymous referee for their valuable suggestions during the peer-review process.
This research has made use of NASA's Astrophysics Data System (ADS) bibliographic services. 
We acknowledge the community efforts devoted to the development of the following open-source packages that were used in this work: numpy (\href{http:\\numpy.org}{numpy.org}), matplotlib (\href{http:\\matplotlib.org}{matplotlib.org}), and astropy (\href{http:\\astropy.org}{astropy.org}).

\end{acknowledgements}

\bibliographystyle{aa}
\bibliography{ref}

\end{document}